\documentclass[12pt]{article}
\usepackage{graphicx}

\newcommand\pubnumber{TUM-HEP-784/10}
\newcommand\pubdate{\today}

\def\napoli{Physik Department, Technische Universit\"at M\"unchen,
James-Franck-Stra{\ss}e, \\D-85748 Garching, Germany}
\def\support{\footnote{This work was partially supported by GRK 1054 of Deutsche Forschungsgemeinschaft.}}

\def\Title#1{\begin{center} {\Large #1 } \end{center}}
\def\Author#1{\begin{center}{ \sc #1} \end{center}}
\def\Address#1{\begin{center}{ \it #1} \end{center}}

\newcommand\pubblock{\rightline{\begin{tabular}{l} \pubnumber\\
         \pubdate  \end{tabular}}}
\newenvironment{Abstract}{\begin{quotation}  }{\end{quotation}}
\newenvironment{Presented}{\begin{quotation} \begin{center} 
             PRESENTED AT\end{center}\bigskip 
      \begin{center}\begin{large}}{\end{large}\end{center} \end{quotation}}
\def\Acknowledgements{\bigskip  \bigskip \begin{center} \begin{large}
             \bf ACKNOWLEDGEMENTS \end{large}\end{center}}

\def\beq{\begin{equation}}
\def\eeq#1{\label{#1}\end{equation}}
\def\eeqn{\end{equation}}

\def\beqa{\begin{eqnarray}}
\def\eeqa#1{\label{#1}\end{eqnarray}}
\def\eeqan{\end{eqnarray}}

\let\bar=\overbar

\def\Dslash{\not{\hbox{\kern-4pt $D$}}}
\def\dslash{\not{\hbox{\kern-2pt $\del$}}}

\def\msb{{\bar{\ssstyle M \kern -1pt S}}}

\begin{document}

\begin{titlepage}
\pubblock

\vfill
\Title{Flavour physics with a 4th generation}
\vfill
\Author{Tillmann Heidsieck\support}
\Address{\napoli}
\vfill
\begin{Abstract}
We present an overview of recent work on flavour physics in the presence of a sequential fourth
generation. We will discuss shortly the constraints on the new parameters and in the reminder
present predictions for observables like ${\rm Br}(B_s\rightarrow \mu^+\mu^-)$, ${\rm Br}(K\rightarrow \pi\nu\bar\nu)$
and the indirect CP violation $S_{\psi\phi}$ in the $B_s$ system. We will further stress
the importance of $\varepsilon^\prime/\varepsilon$ as a possible constraint once reliable
lattice results for $B_6$ and $B_8$ become available. Lepton flavour violation is also briefly discussed
in view of prospects for $\tau$ physics at an upgraded flavour factory as well as upcoming experiments
for $\mu\rightarrow e\gamma$ and $\mu-e$ conversion in nuclei.
\end{Abstract}
\vfill
\begin{Presented}
6th International Workshop on the CKM Unitarity Triangle \\
Warwick, England,  September 6th - 10th, 2010
\end{Presented}
\vfill
\end{titlepage}
\def\thefootnote{\fnsymbol{footnote}}
\setcounter{footnote}{0}

\section{Introduction}
One of the most simple extensions of the Standard Model (SM3) is the addition of a
sequential fourth generation (4G). This model is one the one hand highly restricted
by near three generation unitarity as well as electroweak precision tests~\cite{Eberhardt:2010bm,Chanowitz:2009mz,Erler:2010sk} 
and $\Delta F=2$ observables~\cite{Bobrowski:2009ng,Soni:2010xh,Buras:2010pi}. On the other hand still leaves room for 
sometimes huge effects in soon to be measured observables.
The generalisation of the CKM matrix to four generations yields five new parameters $\theta_{14},\theta_{24},\theta_{34},\delta_{14},\delta_{24}\,.$
Together with the two new quark masses this gives a total of $7$ new parameters in the quark sector. In the lepton
sector the new mixing angles are highly restricted through
an interplay of bounds from rare $\mu$ and $\tau$ decays and their measured lifetimes~\cite{Lacker:2010zz,Buras:2010cp}.
The recent efforts in studying flavour physics in the presence of a fourth generation~\cite{Bobrowski:2009ng,Hou:2005yb,Hou:2006mx,Arhrib:2006pm,Soni:2008bc,Soni:2010xh,Buras:2010pi,Buras:2010nd,Buras:2010cp}
show clearly that in spite of its few new parameters the SM4 can not be excluded yet.
\begin{figure}[ht]
\begin{center}
\includegraphics[width=.48\textwidth]{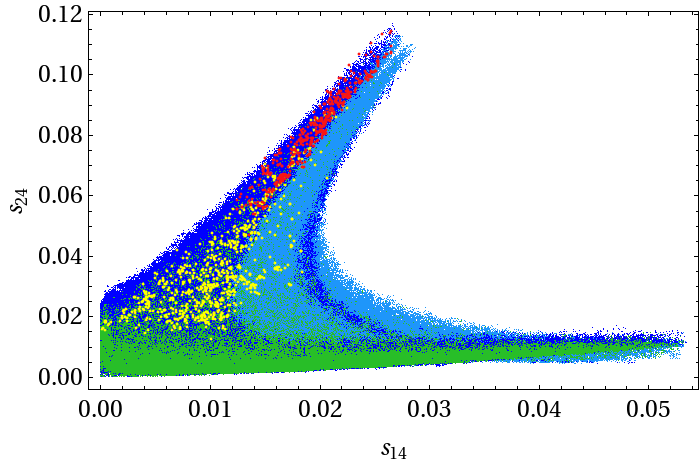}\hspace{.03\textwidth}
\includegraphics[width=.48\textwidth]{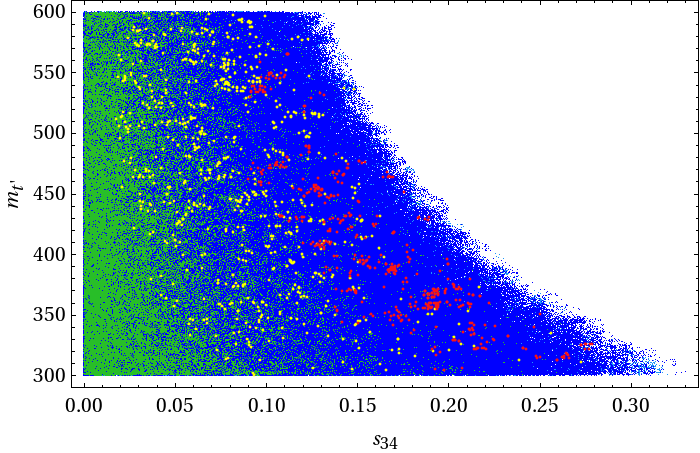}
\end{center}
\vspace{-.5cm}
\caption{$s_{24}$ vs $s_{14}$ (left panel) and $m_{t^\prime}$ as a function of $s_{34}$ (right panel) both show how constrained the
new parameters are.\label{fig:constraints}}
\end{figure}
In figure \ref{fig:constraints} we show on the left panel the correlation of $s_{14}$ and $s_{24}$, this strong correlation is
due to the directly measured CKM elements, Kaon physics and electroweak precision measurements~\cite{Eberhardt:2010bm,Buras:2010pi}. On the right panel of figure \ref{fig:constraints}
we show $m_{t^\prime}$ as a function of $s_{34}$ this correlation can be directly linked to the constraint coming from the $T$ parameter~\cite{Eberhardt:2010bm,Chanowitz:2009mz,Erler:2010sk}.
For the global analysis we use the convenient colour-coding specified below and in table table \ref{tab:Bscenarios}.  The large black point represents the SM3.
Light blue and dark blue  points stand for the results of our global analysis of the SM4 
with the following distinction: light blue stands for  ${\rm Br}(K_L\rightarrow \pi^0\nu\bar\nu) > 2\cdot 10^{-10}$
and dark blue for ${\rm Br}(K_L\rightarrow \pi^0\nu\bar\nu) \leq 2\cdot 10^{-10}$. 
\begin{table}[ht]
\begin{center}
\begin{tabular}{c|ccc}
 		&BS1 (yellow) &BS2 (green)	& BS3 (red) 	\\ \hline
$S_{\psi\phi}$	& $0.04\pm 0.01$& $0.04\pm 0.01$ & $ \geq 0.4$ 	\\ 
${\rm Br}(B_s\to\mu^+\mu^-)$	& $(2\pm 0.2)\cdot 10^{-9}$ & $(3.2\pm 0.2)\cdot 10^{-9} $   &   $\geq 6\cdot 10^{-9}$\\ \hline
\end{tabular}
\caption{Three scenarios for  $S_{\psi\phi}$ and  ${\rm Br}(B_s\to\mu^+\mu^-)$.} \label{tab:Bscenarios}
\end{center}
\end{table}

\section{Rare Decays and CP violation}
\begin{figure}[ht]
\begin{center}
\includegraphics[width=.48\textwidth]{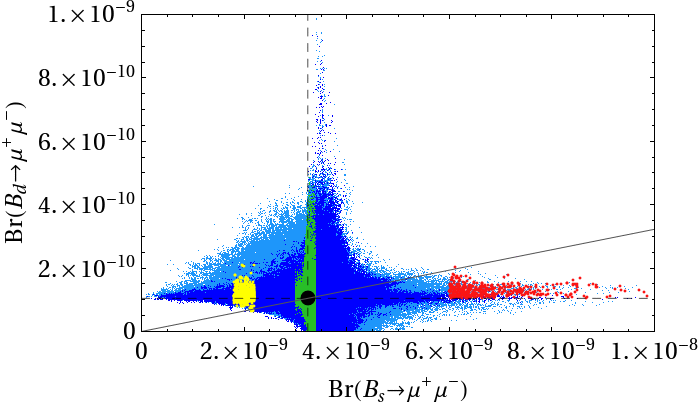}\hspace{.03\textwidth}
\includegraphics[width=.48\textwidth]{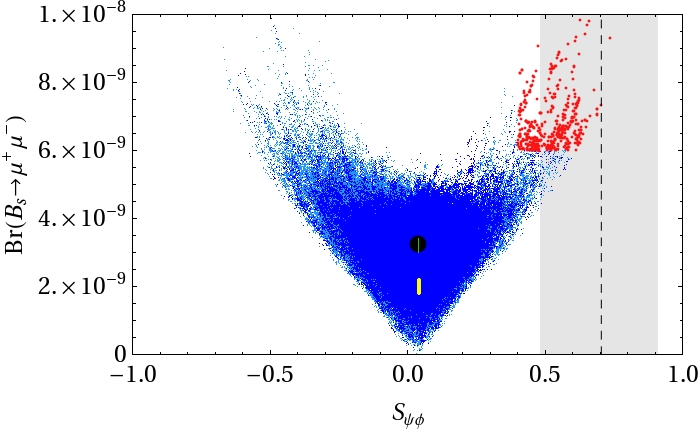}
\end{center}
\vspace{-.5cm}
\caption{${\rm Br}(B_s\to\mu^+\mu^-)$ in correlation with ${\rm Br}(B_d\to\mu^+\mu^-)$ (left panel) and $S_{\psi\phi}$ (right panel). \label{fig:Bsmumu}}
\end{figure}

\begin{figure}[ht]
\begin{center}
\includegraphics[width=.48\textwidth]{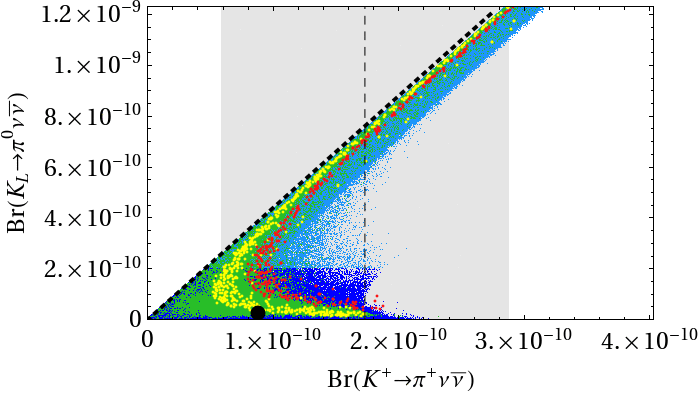}
\end{center}
\vspace{-.5cm}
\caption{${\rm Br}(K_L\to\pi^0\nu\bar\nu)$ as a function of ${\rm Br}(K^+\to\pi^+\nu\bar\nu)$. The dotted line corresponds to the model-independent GN bound.\label{fig:Kpinunu}}
\end{figure}

With the data collected till the end of 2011, the LHC experiments will be able to exclude 
${\rm Br}(B_s\rightarrow \mu^+\mu^-)$ down to nearly the SM3 value. This of course prompts for the analysis of
this decay in the context of models beyond the SM3. In figure \ref{fig:Bsmumu} we show ${\rm Br}(B_s\rightarrow \mu^+\mu^-)$
in correlation with ${\rm Br}(B_d\rightarrow \mu^+\mu^-)$ and $S_{\psi\phi}$. There are two interesting features.
On the left panel we find a anti-correlation between ${\rm Br}(B_s\rightarrow \mu^+\mu^-)$ and ${\rm Br}(B_d\rightarrow \mu^+\mu^-)$,
which allows for any of the both branching ratios to be above or below the SM3 expectation but not simultaneously. On the right panel
we find a strong dependence of ${\rm Br}(B_s\rightarrow \mu^+\mu^-)$ on $S_{\psi\phi}$ in the case of a big $S_{\psi\phi}$. Additionally
we note that for a suppressed ${\rm Br}(B_s\rightarrow \mu^+\mu^-)$ we expect  $S_{\psi\phi}$ to be SM3 like.
Another upcoming experiment (NA62) intends to measure ${\rm Br}(K^+\rightarrow \pi^+\nu\bar\nu)$ to an accuracy of roughly $10\%$.
In figure \ref{fig:Kpinunu} we show ${\rm Br}(K_L\to\pi^0\nu\bar\nu)$ as a function of ${\rm Br}(K^+\to\pi^+\nu\bar\nu)$. ${\rm Br}(K_L\to\pi^0\nu\bar\nu)$ 
can be enhanced by orders of magnitude above the SM3 expectation which implies an automatic enhancement of ${\rm Br}(K^+\to\pi^+\nu\bar\nu)$.
However the reverse is not true, ${\rm Br}(K^+\to\pi^+\nu\bar\nu)$ can be enhanced while ${\rm Br}(K_L\to\pi^0\nu\bar\nu)$ stays
at its SM3 value or even below. The cut in ${\rm Br}(K^+\to\pi^+\nu\bar\nu)$ on the lower axis is due to ${\rm Br}(K_L\rightarrow \mu^+\mu^-)_{\rm SD}$.

\boldmath
\subsection{The importance of $\varepsilon^\prime/\varepsilon$}
\unboldmath
The ratio of direct over indirect CP violation the neutral Kaon decays is well measured but theoretical predictions suffer from
hadronic uncertainties parametrised by $R_6$ and $R_8$. 
To address the issue of the two unknown parameters we chose four different scenarios for $(R_6,R_8)$
and studied the effect $\varepsilon^\prime/\varepsilon$ could have as a constraint on the SM4.
\begin{figure}[ht]
\begin{center}
\includegraphics[width=.48\textwidth]{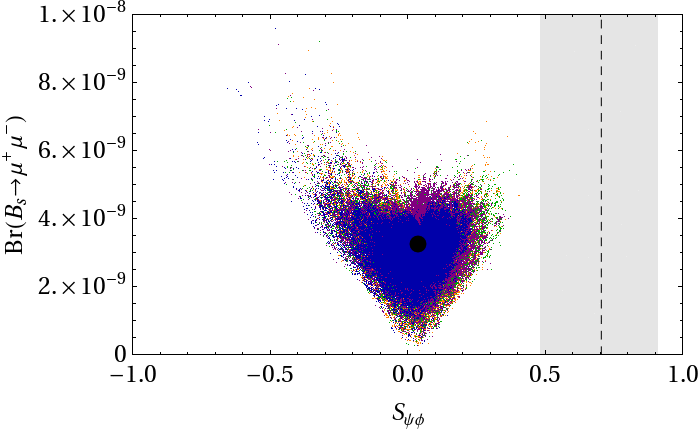}\hspace{.03\textwidth}
\includegraphics[width=.48\textwidth]{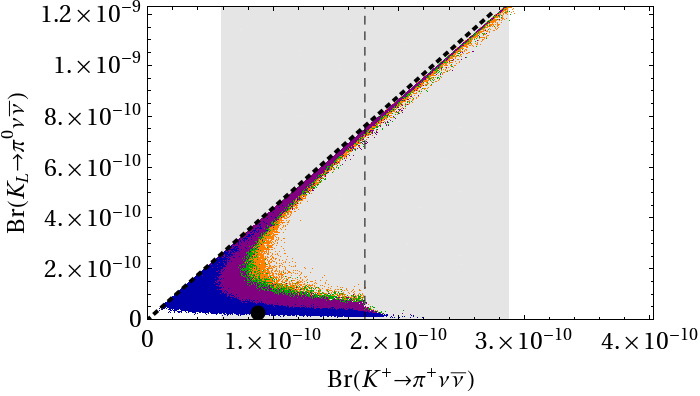}
\caption{The correlation ${\rm Br}(B_s\to\mu^+\mu^-)$ vs. $S_{\psi\phi}$ (left panel) ${\rm Br}(K_L\to\pi^0\nu\bar\nu)$ as a function of ${\rm Br}(K^+\to\pi^+\nu\bar\nu)$ (right panel) 
after including the $\varepsilon^\prime/\varepsilon$-constraint (colour-coding according to Tab.~\ref{tab:Rscenarios}).\label{fig:epsconst}}
\end{center}
\end{figure}
\begin{table}[ht] 
\begin{center}
\begin{tabular}{ccc||ccc}
$R_6$	& $R_8$	& & $R_6$ & $R_8$ &	 	\\ \hline
$1.0$	& $1.0$	& dark blue	&  $1.5$	& $0.8$	& purple	\\ 
$2.0$	& $1.0$	& green &  $1.5$	& $0.5$	& orange	\\ \hline
\end{tabular}
\caption{Four scenarios for the parameters $R_6$ and $R_8$\label{tab:Rscenarios}}
\end{center}
\end{table}
In figure \ref{fig:epsconst} we show the impact of $\varepsilon^\prime/\varepsilon$ as a constraint on the correlations
${\rm Br}(B_s\to\mu^+\mu^-)$ vs. $S_{\psi\phi}$ and ${\rm Br}(K_L\to\pi^0\nu\bar\nu)$ vs. ${\rm Br}(K^+\to\pi^+\nu\bar\nu)$
for the four different scenarios for the unknown hadronic parameters $R_6$ and $R_8$. Note that for all scenarios the allowed
range for $0<S_{\psi\phi}$ is severely constrained. This behaviour can be easily understood; for $S_{\psi\phi}>0$ the
contributions from the $Z^0$ penguins with $t$ and $t^\prime$ have the same sign and overcompensate the contributions of the QCD penguins .
The solution to this problem is to reduce the effect of the $Z^0$ penguins while increasing the importance of the QCD penguins, e.g. the 'orange' scenario.

\section{Lepton Flavour Violation}
\begin{figure}[ht]
\begin{center}
\includegraphics[width=.48\textwidth]{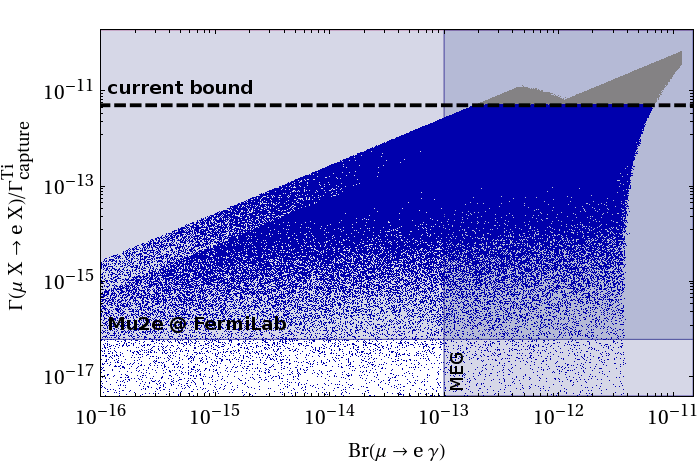}\hspace{.03\textwidth}
\includegraphics[width=0.48 \textwidth]{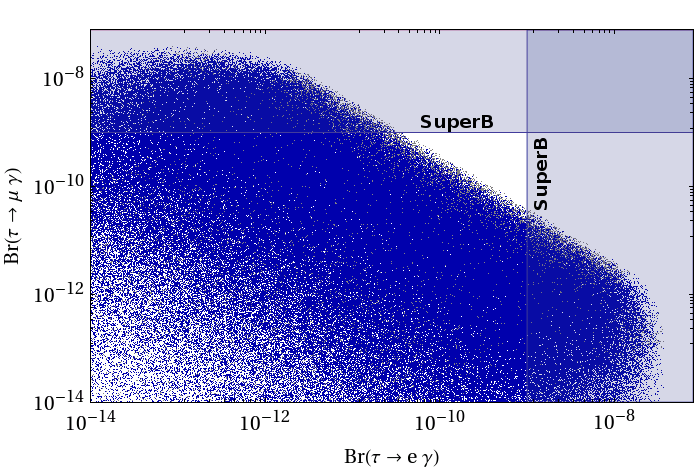}
\end{center}
\vspace{-.5cm}
\caption{Correlation between ${\rm Br}(\mu\rightarrow e\gamma)$ and ${\rm R}(\mu{\rm Ti}\rightarrow e{\rm Ti})$ (left panel). 
Correlation between ${\rm Br}(\tau\rightarrow \mu\gamma)$ and ${\rm Br}(\tau\rightarrow e\gamma)$ (right panel). 
The shaded areas indicate the expected future experimental bounds on both observables. \label{fig:lfv}}
\end{figure}

The mixing between the fourth and the first three lepton generations is stringently constrained through an interplay of
radiative $\mu$ and $\tau$ decays and their respective tree-level decays~\cite{Lacker:2010zz,Buras:2010cp}.
New experiments for the search of the decay $\mu \rightarrow e \gamma$ and $\mu-e$ conversion in nuclei are
in the progress of being build or in a planning stage. On the left panel of figure~\ref{fig:lfv} we show how these future
measurements will constrain the correlation of $\mu\rightarrow e \gamma$ an $\mu-e$ conversion in nuclei. 
Note that the correlation is rather diffuse, this is due to the fact that for $\mu-e$ conversion several contributions can cancel each
other~\cite{Buras:2010cp}. 
On the right panel of figure~\ref{fig:lfv} we show the correlation between ${\rm Br}(\tau\rightarrow \mu\gamma)$ and ${\rm Br}(\tau\rightarrow e\gamma)$ together with
projected exclusion limits from SuperBelle. The shape of the correlation is due to the $|U_{\tau 4}^*U_{e4}|$ dependence for $\tau\rightarrow e\gamma$ and
the $|U_{\tau 4}^*U_{\mu 4}|$ dependence for $\tau\rightarrow \mu\gamma$. Taking into account that $\mu-e$ conversion constrains $|U_{\mu 4}^*U_{e 4}|$ it is
clear that ${\rm Br}(\tau\rightarrow \mu\gamma)$ and ${\rm Br}(\tau\rightarrow e\gamma)$ can not be at their maximum simultaneously.

\section{Conclusions}
\begin{itemize}
 \item The branching ratio ${\rm Br}(B_s\to\mu^+\mu^-)$ can be enhanced or suppressed in the SM4. However if $S_{\psi\phi} \gg 0$ as suggested by the Tevatron data was indeed true we would expect an enhancement of ${\rm Br}(B_s\to\mu^+\mu^-)$.
 \item In the $K$ system there is independently of the $B$ system much room for in some cases huge effects, however they are correlated among each other.
 \item $\varepsilon^\prime/\varepsilon$ can pose a very stringent constraint on the SM4 if the non-pert. parameters $B_6$ and $B_8$ were known to a decent accuracy.
 \item The new mixing angles in the lepton sector are tightly constrained through upper bounds on rare decays and tree level decays of $\mu$ and $\tau$.
 Still big enhancements of rare decays are possible though not always simultaneously.
\end{itemize}

\Acknowledgements
I want to thank Andrzej J. Buras, Bj\"orn Duling, Christoph Promberger, Thorsten Feldmann and Stefan Recksiegel 
for the fruitful collaboration. This work was partially supported by GRK 1054 of Deutsche Forschungsgemeinschaft.

\end{document}